\documentclass[conference]{IEEEtran}
\IEEEoverridecommandlockouts
\usepackage{cite}
\usepackage{amsmath,amssymb,amsfonts}
\usepackage{algorithmic}
\usepackage{graphicx}
\usepackage{textcomp}
\usepackage{xcolor}
\usepackage{soul}
\usepackage{comment}
\def\BibTeX{{\rm B\kern-.05em{\sc i\kern-.025em b}\kern-.08em
    T\kern-.1667em\lower.7ex\hbox{E}\kern-.125emX}}
\begin{document}

\title{Multi-channel Medium Access Control Protocols for Wireless Networks within Computing Packages\vspace{-0.2cm}\thanks{This work is supported by the European Commission under H2020 grant WiPLASH (GA 863337) and HE grant WINC (GA 101042080).}
}

\author{\IEEEauthorblockN{Bernat Oll\'e\IEEEauthorrefmark{1},
Pau Talarn\IEEEauthorrefmark{1},
Albert Cabellos-Aparicio\IEEEauthorrefmark{1}, 
Filip Lemic\IEEEauthorrefmark{2},
Eduard Alarc\'on\IEEEauthorrefmark{1}, and
Sergi Abadal\IEEEauthorrefmark{1}}
\IEEEauthorblockA{\IEEEauthorrefmark{1}NaNoNetworking Center in Catalunya (N3Cat), Universitat Polit\`ecnica de Catalunya, 08034 Barcelona, Spain}
\IEEEauthorblockA{\IEEEauthorrefmark{2}AI-driven Systems Lab, i2Cat Foundation, 08034 Barcelona, Spain}\vspace{-0.5cm}
}

\maketitle

\begin{abstract}
Wireless communications at the chip scale emerge as a interesting complement to traditional wire-based approaches thanks to their low latency, inherent broadcast nature, and capacity to bypass pin constraints. However, as current trends push towards massive and bandwidth-hungry processor architectures, there is a need for wireless chip-scale networks that exploit and share as many channels as possible. 
In this context, this work addresses the issue of channel sharing by exploring the design space of multi-channel Medium Access Control (MAC) protocols for chip-scale networks. Distinct channel assignment strategies for both random access and token passing are presented and evaluated under realistic traffic patterns. It is shown that, even with the improvements enabled by the multiple channels, both protocols maintain their intrinsic advantages and disadvantages.
\end{abstract}
\vspace{-0.1cm}


\section{Introduction}
Efficient integrated networks at the chip scale within Systems-in-Package (SiPs) 
are a prerequisite for high performance in such computing systems. Currently, most systems incorporate a Network-in-Package (NiP) consisting of a set of on-chip routers and intra-/inter-chip wired links \cite{Marculescu2009, Bertozzi2014}. However, recent scaling \cite{salahuddin2018era, mayr2019spinnaker}, specialization \cite{krishna2020data, wang2022application}, and disintegration trends \cite{kannan2016exploiting,naffziger2021pioneering} are increasing the pressure placed on the interconnect, to the point that new communication paradigms may be required \cite{karkar2016survey, ganguly2022interconnects}. 

\begin{figure}[!t] 
\centering
\includegraphics[width=0.45\columnwidth]{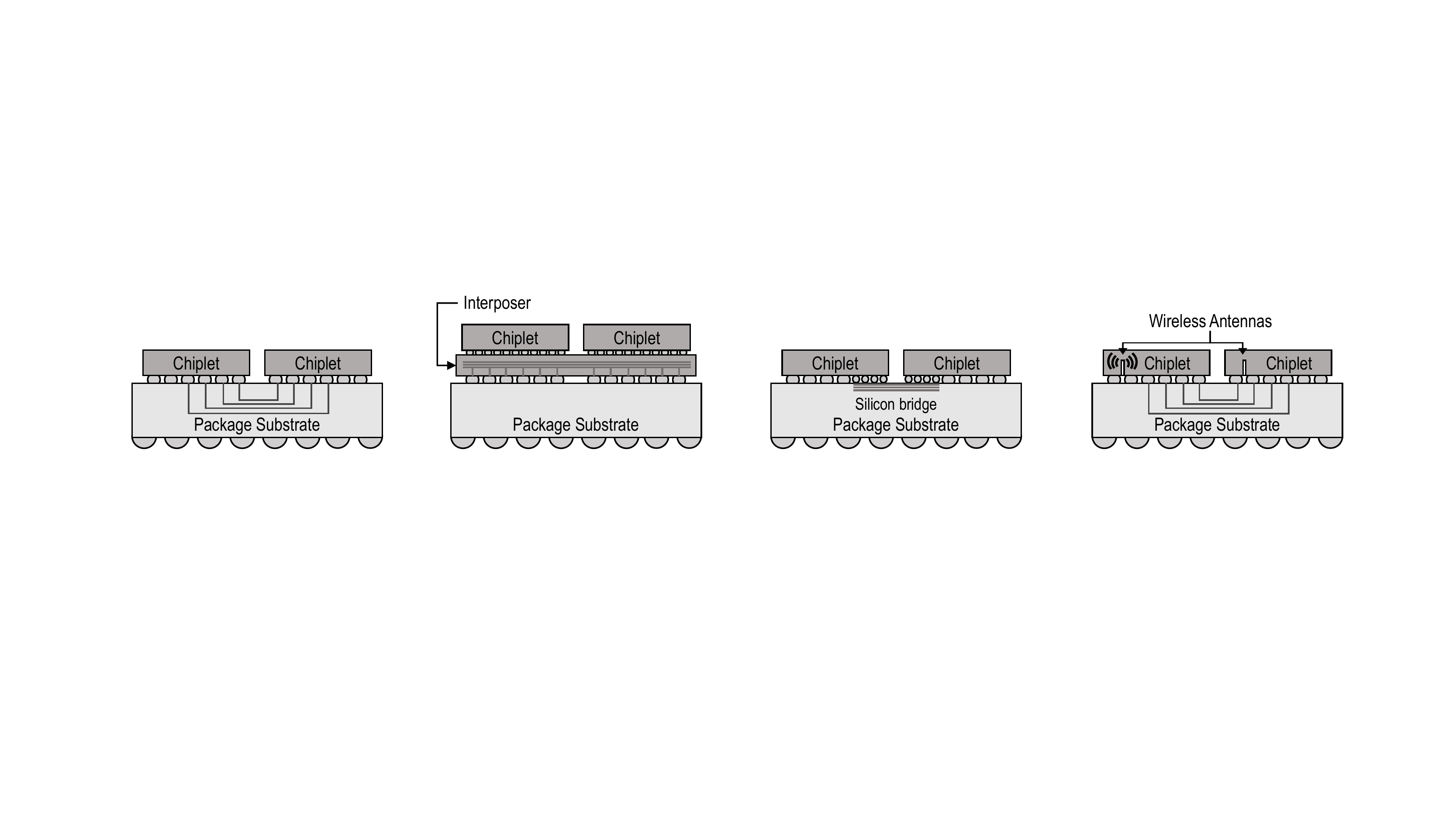}
\vspace{-0.25cm}
\caption{Pictorial view of a wireless chip-to-chip communication link.} \label{wichip}
\vspace{-0.4cm}
\end{figure}

Among the emerging alternatives, wireless chip-scale communications stand as a promising contender \cite{matolak2012wireless, laha2014new, abadal2022graphene, narde2019intra}. This communication paradigm relies the use of modulated electromagnetic waves for data transmission using the chip package as communications medium (Fig.~\ref{wichip}). The resulting \textit{wireless in-package links} provide low latency, inherent broadcast capabilities, and global reconfigurability. 

Since the communications medium is shared, wireless in-package communications require Medium Access Control (MAC) protocols to avoid or manage wasteful collisions. In this scenario, MAC protocols generally reduce to variants of multiplexing, random access, or token passing \cite{ditomaso2015winoc, mansoor2015reconfigurable, Duraisamy2015, Mestres2016, jog2021one, franques2021fuzzy, ouyang2021architecting}. Even though recent works have demonstrated that computing packages could support a few frequency \cite{Rayess2017,timoneda2018channel} and space channels \cite{Baniya2018,rodriguez2022towards}, it is still unclear how MAC protocols can benefit from them. This is because more than a few channels are needed to implement truly scalable frequency/space multiplexing techniques \cite{ditomaso2015winoc}, and most importantly, because multi-channel variants of random access and token passing have not been explored 
yet.



This paper aims to bridge this gap by focusing on the study of multi-channel versions of the two most representative protocol types in chip-scale scenarios, i.e. random access and token passing. In particular, the main contributions are as follows. We first describe the different ways we can extend random access and token passing with a small set of channels in Sec.~\ref{mac_protocols}. Then, in Sec.~\ref{sec:MACmodels}, we evaluate these protocol variants with traffic models typically used to mimic multiprocessor workloads \cite{soteriou2006statistical}. 
This analysis sheds light on the impact of channel assignment on the protocol performance, as summarized in Sec.~\ref{sec:discussion} and concluded in Sec.~\ref{sec:conclusions}.

\section{Multi-channel MAC Protocols}
\label{mac_protocols}
In this work, we describe three 
distinct channel assignment strategies for random access and token passing. As baselines, we take BRS \cite{Mestres2016} for random access and the baseline from \cite{franques2021fuzzy} for token passing. The strategies presented here are not provably optimal, but they are simple (as required by the resource constraints of the chip-scale scenario) and representative of the potential techniques that can be used.

\subsection{Assignment Methods for BRS}
In random access protocols such as BRS \cite{Mestres2016}, nodes contend for channel access and back off if the channel is busy or there is a collision. Assuming $N$ nodes, we study three ways to reduce the collision probability using $N_{c}$ channels, namely:


\vspace{0.1cm}\noindent
\textbf{AS1:} Channels are assigned to nodes individually and randomly. When a node has a packet to transmit, the node is assigned a random channel. If the channel is  busy or there is a collision, nodes undergo a random back off and also choose a random channel to use in the next attempt.

\begin{figure}[!t] 
\centering
\includegraphics[width=1.45in]{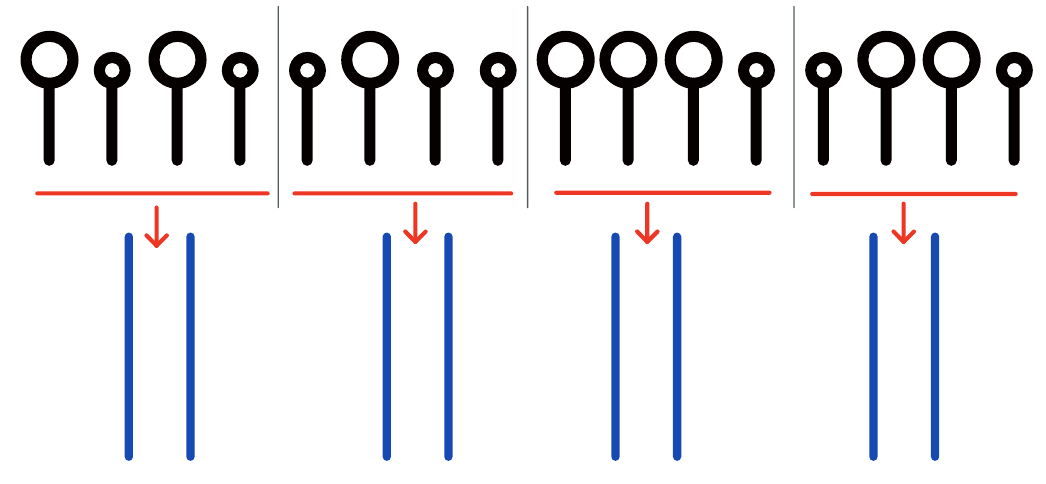}
\includegraphics[width=0.4in]{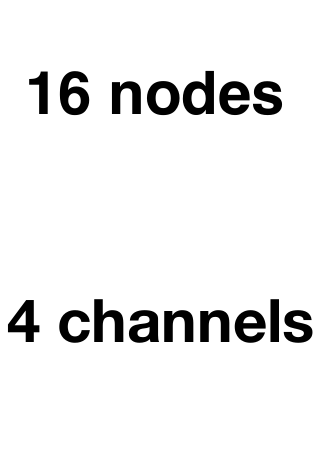}
\includegraphics[width=1.45in]{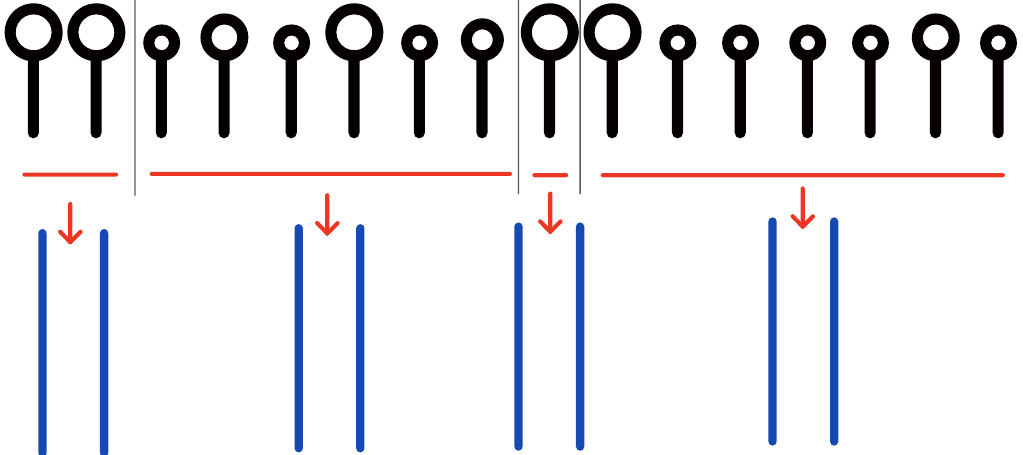}
\vspace{-0.2cm}
\caption{Graphical representations of assignment techniques AS2 (left) and AS3 (right) for BRS assuming 16 nodes and 4 channels.} \label{MULTIBRS}\label{MULTIBRSP} 
\vspace{-0.3cm}
\end{figure}

\vspace{0.1cm}\noindent
\textbf{AS2:} Each channel is assigned to $\tfrac{N}{N_{c}}$ nodes statically following a uniform distribution, this is, assuming that all nodes have the same load (see Fig.~\ref{MULTIBRS}, left). While this is not optimal for spatially unbalanced traffic, it serves as a baseline.

\vspace{0.1cm}\noindent
\textbf{AS3:} Channels are assigned to a variable number of nodes following a distribution that balances the load in each channel (see Fig.~\ref{MULTIBRS}, right). To that end, nodes are ordered based on the expected normalized load and assigned to each channel in order following a greedy algorithm.

\subsection{Assignment Methods for Token Passing}
In token passing \cite{holland2001rate}, typically, all $N$ nodes are sorted forming a virtual ring and the token is passed in order through that ring. In a version with $N_{c}$ channels, each channel can be a token. The design decisions then lie on the number of rings and the nodes that form each ring. 
For instance:

\vspace{0.1cm}\noindent
\textbf{AS1:} We assume as many rings as there are channels and map nodes uniformly to each ring. In other words, we distribute them in rings of $\tfrac{N}{N_{c}}$ nodes, regardless of their expected load.  

\vspace{0.1cm}\noindent
\textbf{AS2:} We assume a single virtual ring with multiple tokens circulating in it. In this case, tokens can jump over other tokens: when node $i$ holds a token for multiple cycles during a transmission, idle tokens that arrive at $i$-1 can jump to $i$+1.

\vspace{0.1cm}\noindent
\textbf{AS3:} This strategy is similar to AS1, but nodes are mapped to rings based on their expected load. This may lead to rings of different sizes, but similar in the expected overall load.


\begin{figure}[!t] 
\centering
\includegraphics[width=0.9\columnwidth]{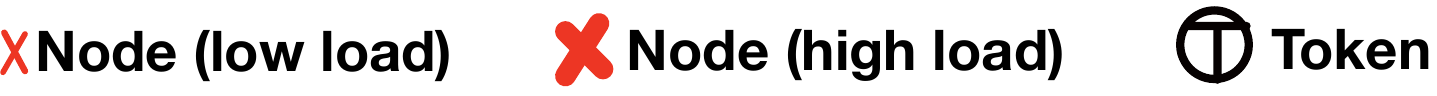}
\vspace{-0.3cm}
\includegraphics[width=0.9in]{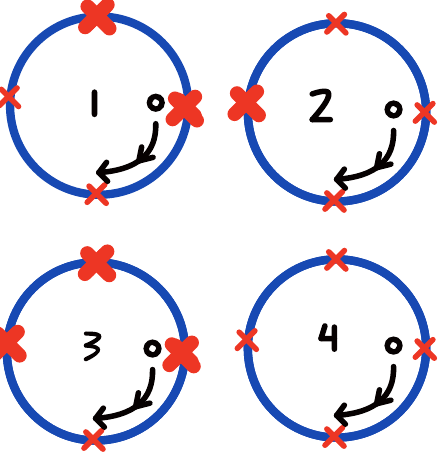}
\hspace{0.2cm}
\includegraphics[width=1.in]{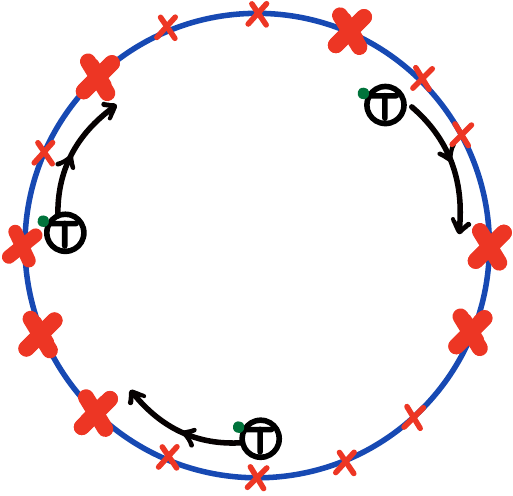}
\hspace{0.2cm}
\includegraphics[width=0.85in]{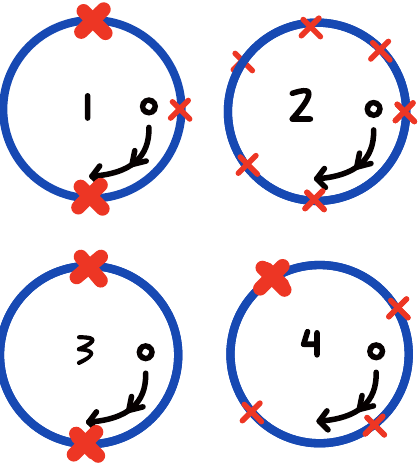}
\caption{Graphical representations of the different assignment techniques for token passing assuming 16 nodes and 4 channels.}\label{MultiRingP}
\vspace{-0.4cm}
\end{figure}

\section{Performance Evaluation}
\label{sec:MACmodels}
The architecture and application parameters are summarized in Table \ref{tab:params}. We implement both single-channel baselines and multi-channel versions of BRS and token passing as finite state machines in a modified version of Multi2sim that models wireless links and supports collision detection \cite{Ubal2012}. The protocols are stressed with synthetic traffic modeling uneven injection distributions (through the $\sigma$ parameter) and bursty temporal behavior (through the Hurst exponent $H$) \cite{soteriou2006statistical}. The default values for the different parameters are $N=64$ nodes, $N_{c}=4$ channels, $H=0.5$ and $\sigma = 1$. Simulations are cycle-accurate.


In all cases, we compare the packet latency (in cycles) and throughput (in packets/cycle) of the different options. Given the high number of protocol strategies and traffic types, instead of plotting the classical latency--throughput curve, we make use of box plots that summarize the latency and throughput statistics. In our plots, the X axis shows the parameters under study. 
The plots have two Y axis: the left axis represents the latency and corresponds to the box plot values, whereas the right axis represents the throughput and corresponds to single-value markers of saturation throughput. 
Since a single packet takes 4 cycles in a single channel to be transmitted, the maximum throughput is 0.25 packets/cycle/channel.

\begin{table}[!t]
            \caption{Characteristics of simulated protocols and applications.}
            \label{tab:traffic2}\label{tab:params}
            \centering
            \vspace{-0.2cm}
            \begin{tabular}{| l | l |} 
            \hline
            Application & Synthetic traffic, H=0.5--0.85, $\sigma$=0.05--100 \\
            System & $N$=64--512 cores, one antenna/core, 1-GHz clock \\
            Network & 80-bit packets (preamble: 20 bits), $N_{c}$=1--4 channels \\ 
            Link & BRS \cite{Mestres2016}, Token passing \\ 
            Physical & On-Off Keying, 20 Gb/s \\ \hline
                
            \end{tabular}
             \vspace{-0.4cm}
        \end{table}
        




\subsection{Number of Channels}
\begin{figure}[!t] 
\centering
\includegraphics[width=0.75\columnwidth]{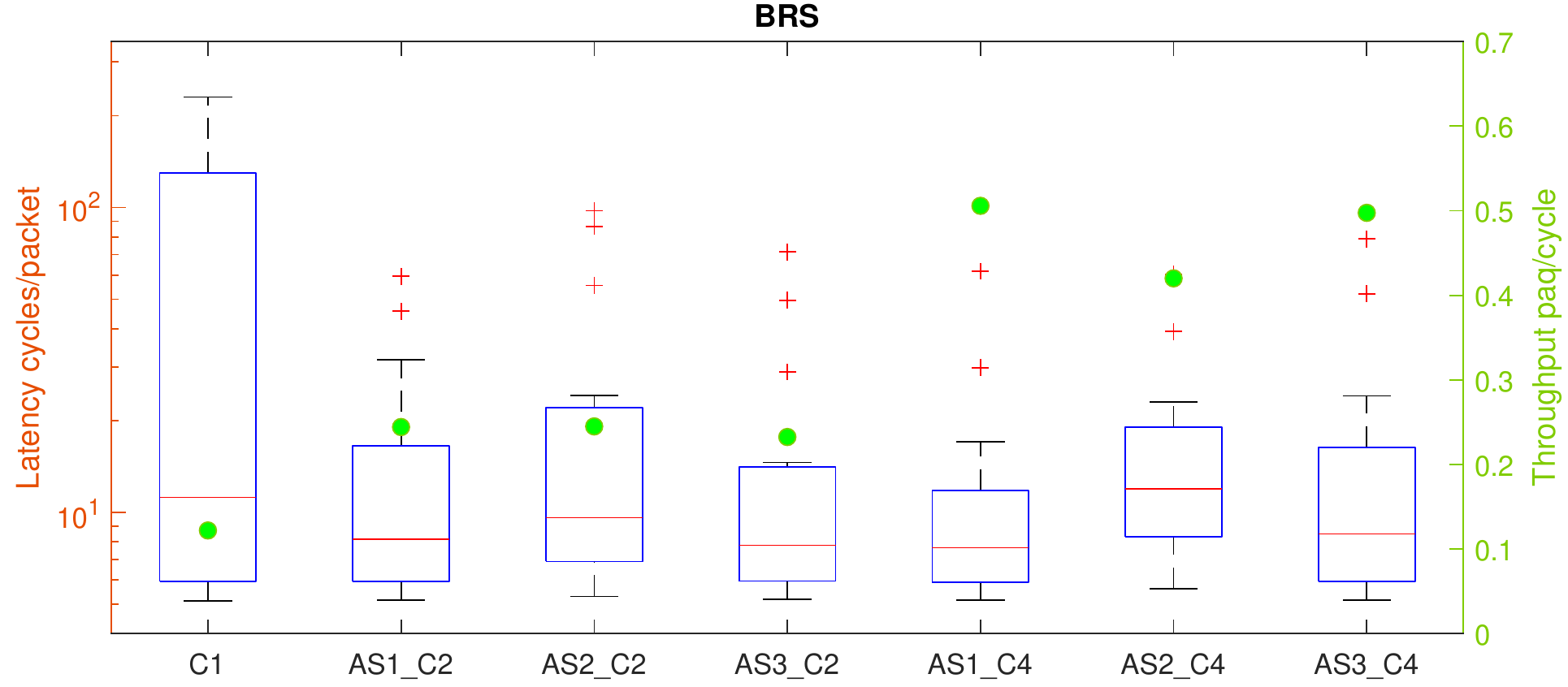}
\includegraphics[width=0.75\columnwidth]{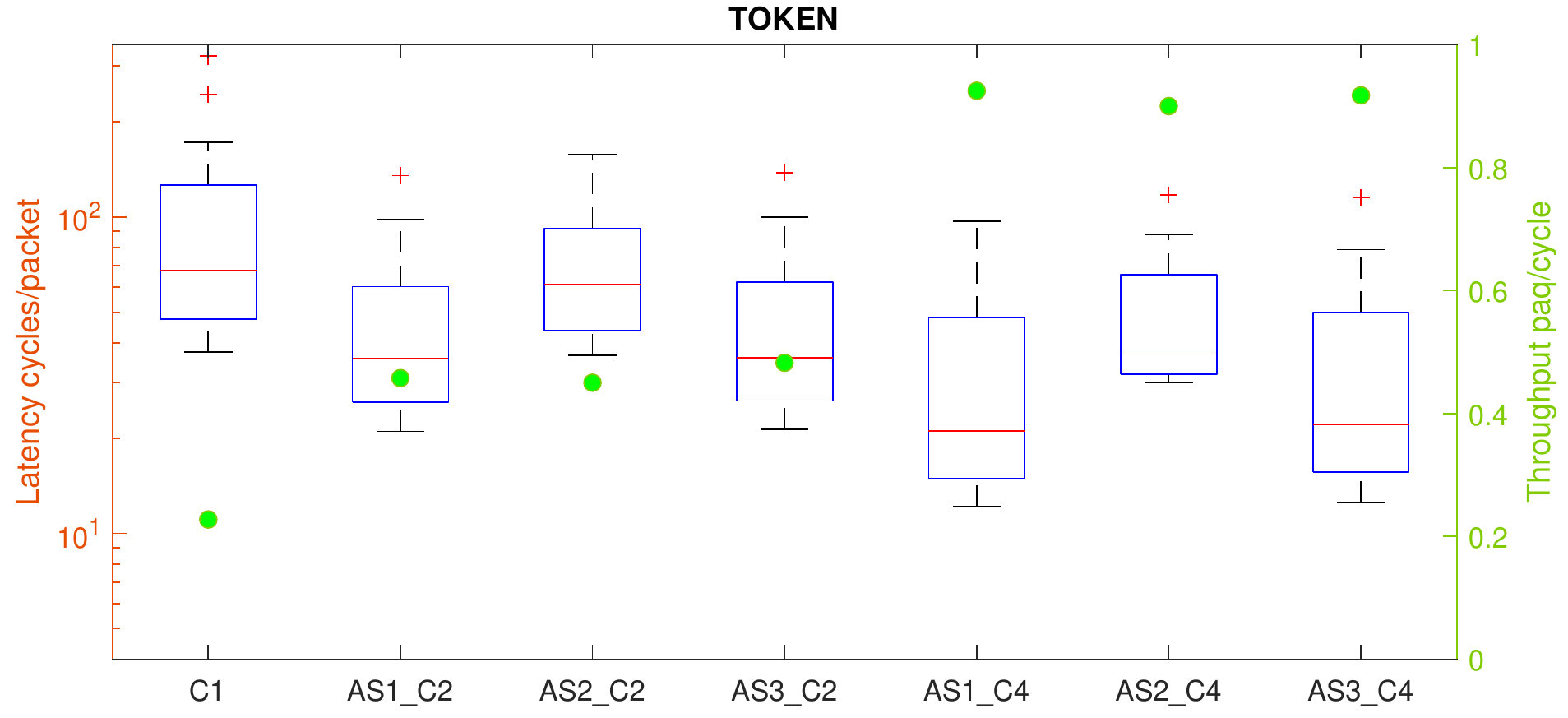}
\vspace{-0.3cm}
\caption{Performance of multi-channel BRS (top) and token passing (bottom) for an increasing number of channels, $C1$ to $C4$, and different assignments.}\vspace{-0.4cm}
\label{boxplotTOKENandBRS}
\end{figure}

Here, we discuss the results shown in Fig.~\ref{boxplotTOKENandBRS} for BRS and token passing and an increasing number of channels. 

\vspace{0.1cm}
\noindent
\textbf{Latency.} In general, it can be observed that BRS is less stable than token in terms of latency as the range of values is larger, with a higher number of outlier points. However, BRS has a much better zero-load latency than token since, in BRS, the protocol allows nodes to start transmitting immediately when the channel is sensed idle. This fact also can explain why independently of the parameters evaluated here (assignment, number of channels) the minimum latency is quite similar. The worst-case latency, however, clearly improves when having multiple channels, as the high load is distributed over multiple channels. On the other hand, in token passing, nodes must wait until they possess the token to start transmitting. For this reason, when the number of nodes is large, $N=64$ in this case, the system remains idle much longer. 

\begin{figure*}[!ht] 
\centering
\includegraphics[width=0.3\textwidth]{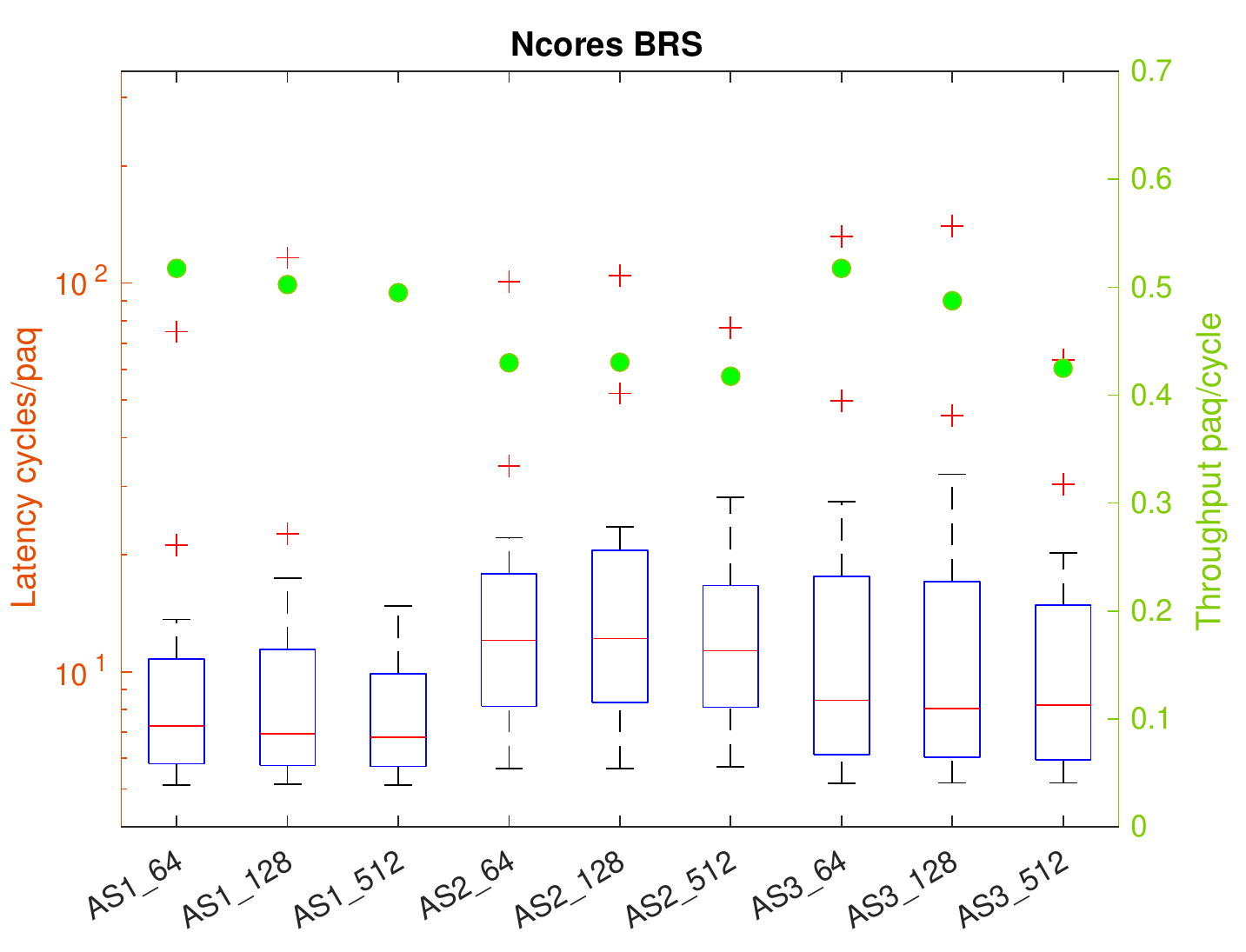}
\includegraphics[width=0.3\textwidth]{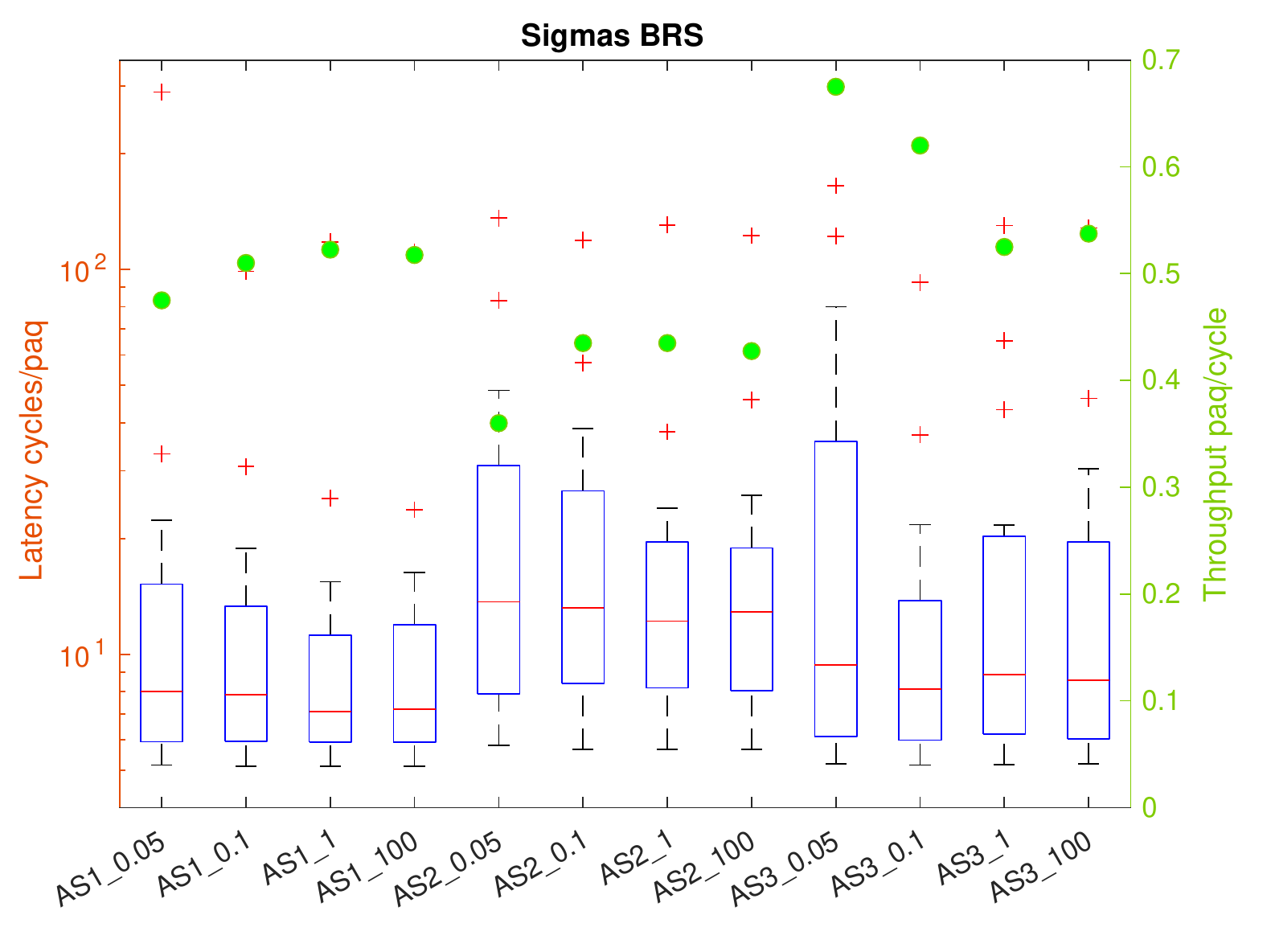}
\includegraphics[width=0.29\textwidth]{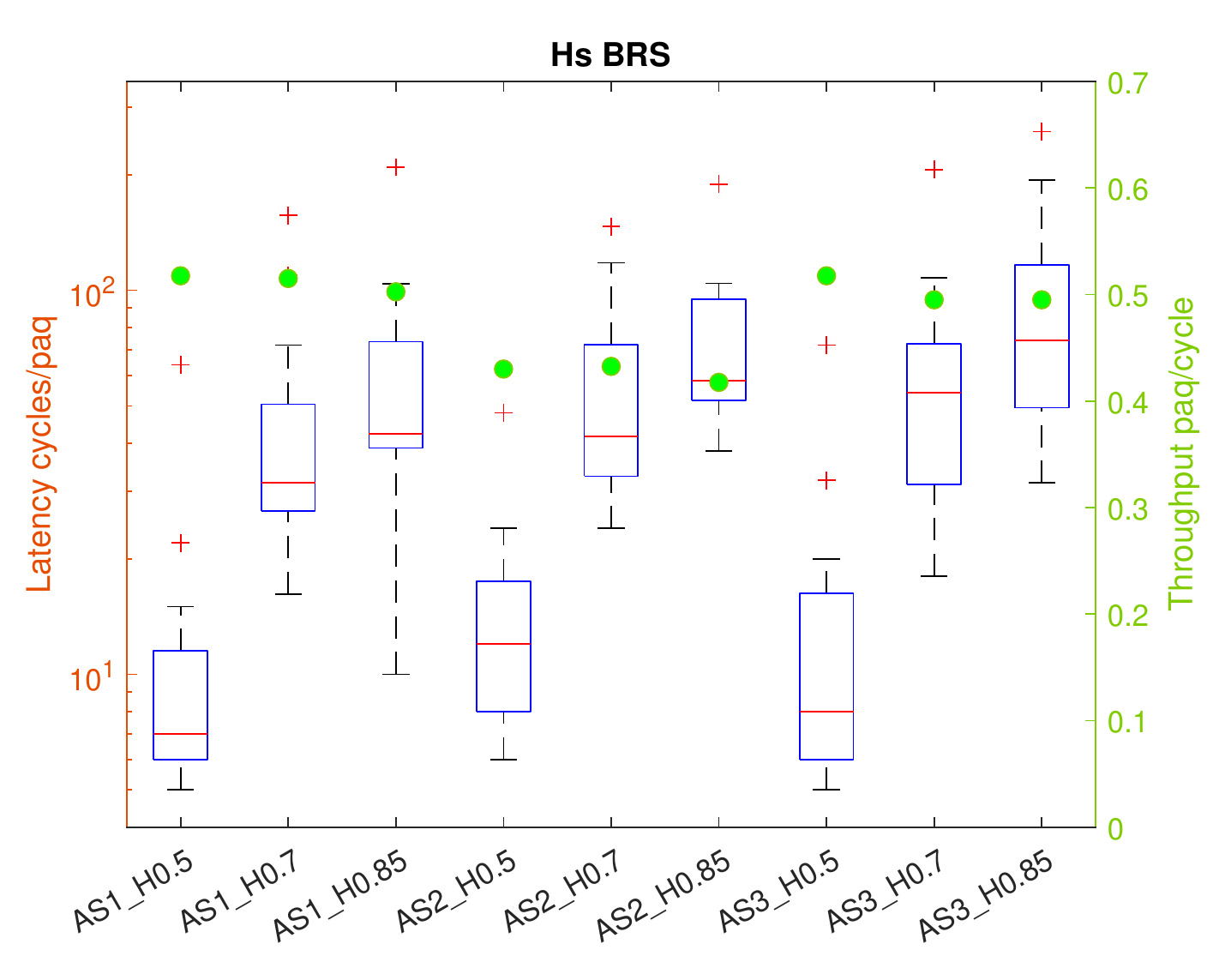}
\vspace{-0.3cm}
\caption{Performance of multi-channel BRS protocol for an increasing number of nodes, $N$=64--512 (left graph), different spatial concentration levels, $\sigma$=0.1--100 (center graph), different temporal burstiness levels, $H$=0.5--0.85 (right graph), and different assignment techniques.}
\label{boxplotBrs}
\vspace{-0.3cm}
\end{figure*}

\begin{figure*}[!ht] 
\centering
\includegraphics[width=0.3\textwidth]{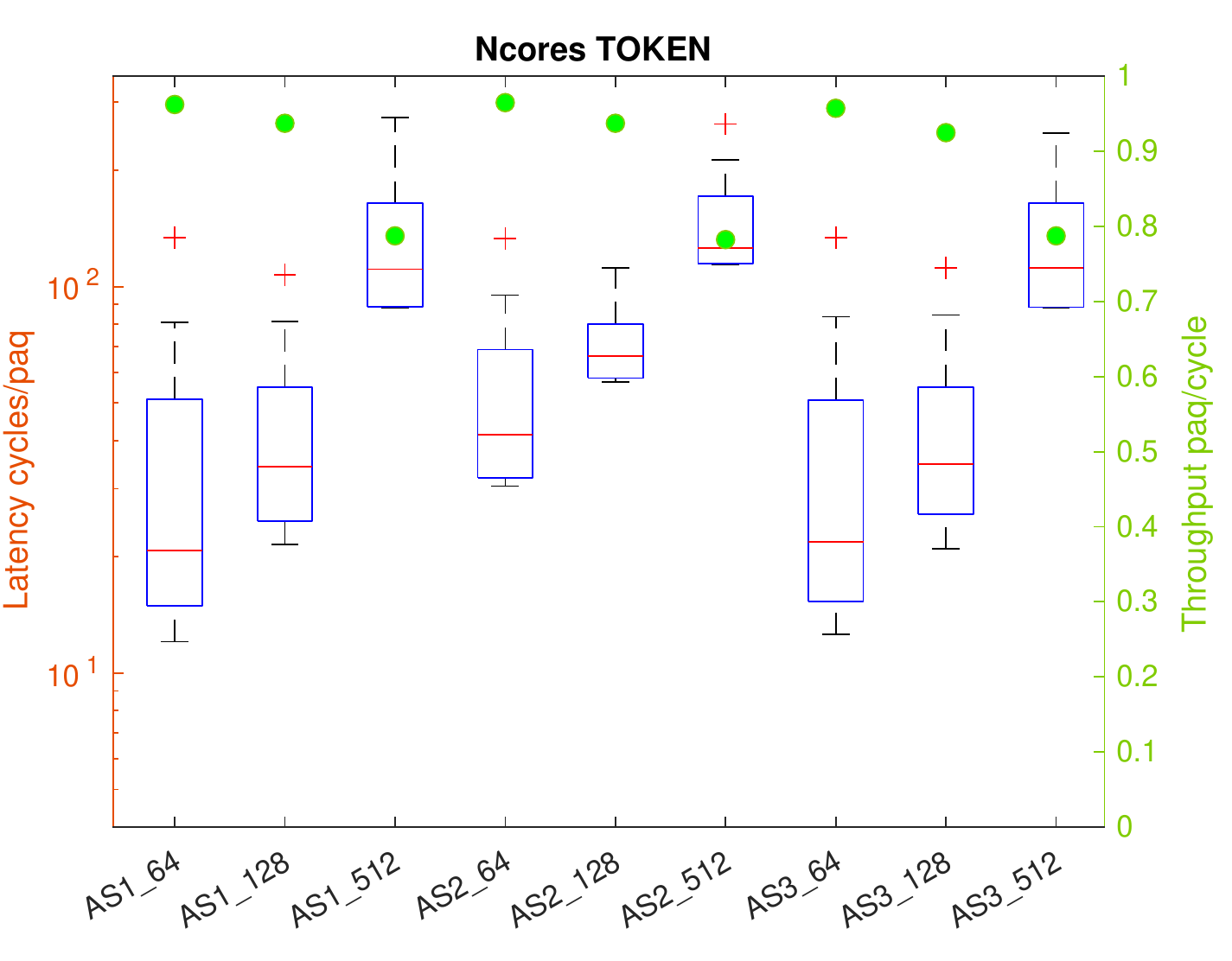}
\includegraphics[width=0.3\textwidth]{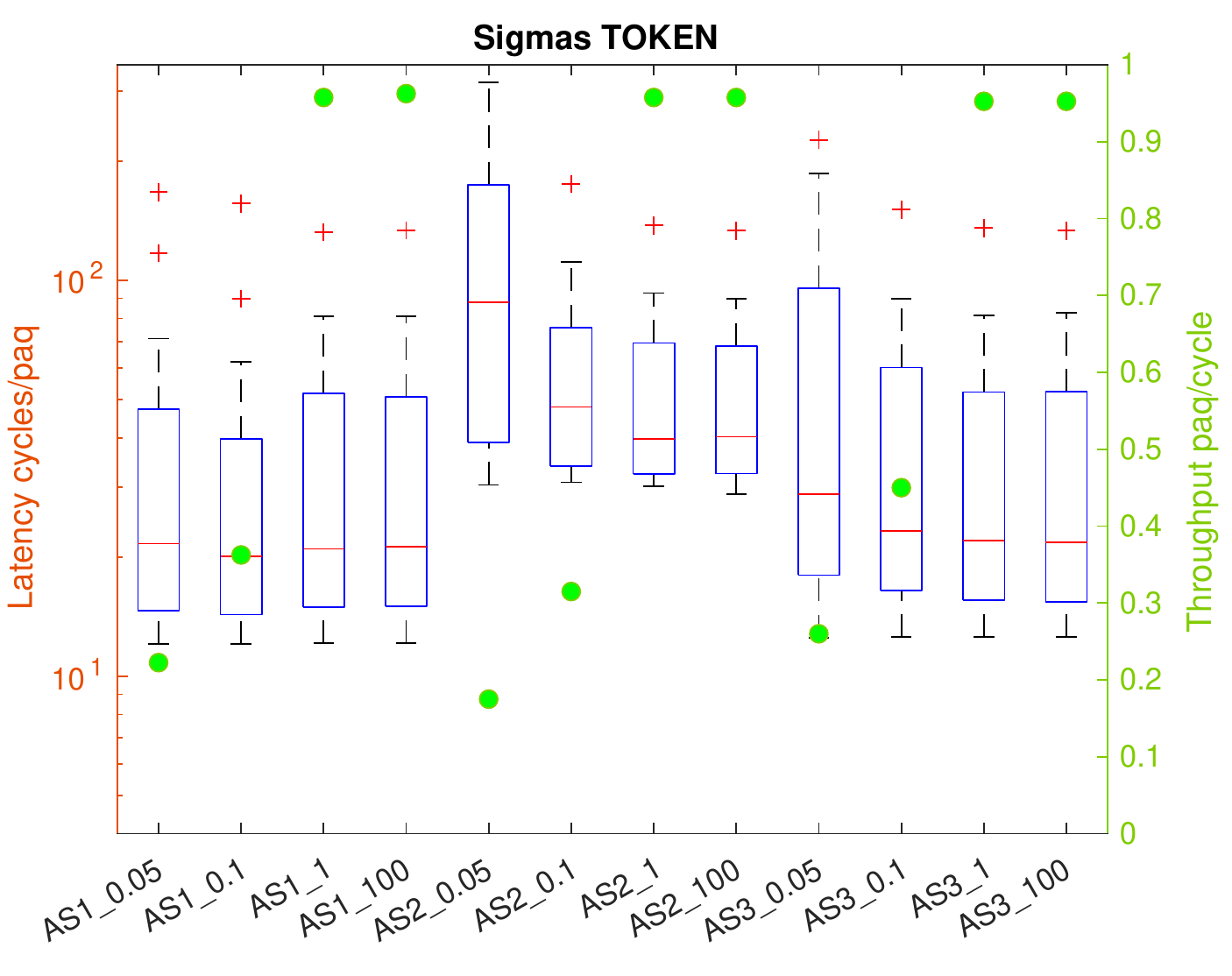}
\includegraphics[width=0.3\textwidth]{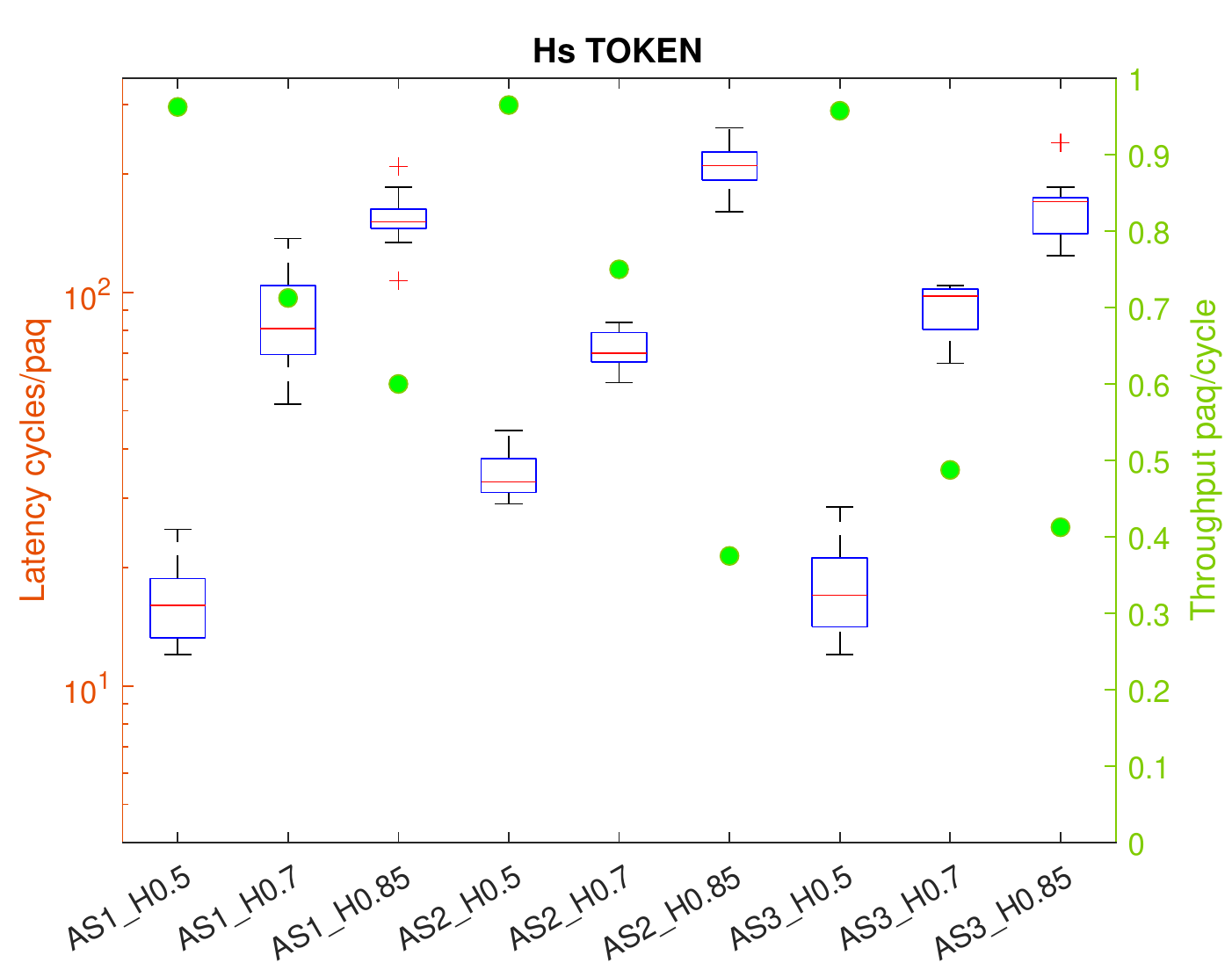}
\vspace{-0.3cm}
\caption{Performance of multi-channel token passing protocol for an increasing number of nodes, $N$=64--512 (left graph), different spatial concentration levels, $\sigma$=0.1--100 (center graph), different temporal burstiness levels, $H$=0.5--0.85 (right graph), and different assignment techniques.}
\label{boxplotToken}
\vspace{-0.4cm}
\end{figure*}

\vspace{0.1cm}
\noindent
\textbf{Throughput.} The results for token passing depict a rather stable increase in saturation throughput as more channels are added, regardless of the assignment method. This could be due to the use, by default, of non-bursty and non-hotspot traffic to evaluate scalability. On the other hand, the results for BRS illustrate a different behavior than in token passing. Firstly, BRS cannot reach a saturation throughput as high as token passing. 
The main reasons are that 
channel contention and multiple collisions lead to channel waste and, hence, to a reduced throughput. 
Furthermore, BRS is more irregular than token passing in terms of saturation throughput as it depends on the percentage of collisions at high loads. As a result, the difference between the saturation throughput achieved for different assignments increases with the number of channels. 

\subsection{Number of Nodes}
Next, we comment on the performance of BRS and token passing for an increasing number of nodes, with $N_{c}=4$. The results are shown in the left charts of Fig.~\ref{boxplotBrs} and Fig.~\ref{boxplotToken}. 


\vspace{0.1cm}
\noindent
\textbf{Latency.} BRS has a much lower latency than token passing due to its ability to transmit when the channel is idle. The span of the latency values differs across number of nodes and assignments, but in general are restrained to similar values because in the end, the same aggregated load ends up being distributed over more nodes. Static assignment of channels (AS2) works worse than the other alternatives.
On the other hand, from the plot of token passing, it is clear that more nodes lead to much higher latency due to the increase of the token turnaround time. In fact, the low-load latency is proportional to the number of nodes in all cases. The span of the latency values is similar across the different system sizes.


\vspace{0.1cm}
\noindent
\textbf{Throughput.} In general, saturation throughput is slightly higher for a lower number of nodes. 
In our protocols, having more nodes means having a higher population and, hence, a higher chance of collisions even for the same load for BRS, and a higher waiting time (or lower probability of having all nodes backlogged) in token passing. It seems, in any case, that BRS is more resilient to the change in the number of nodes as the drop is more subtle, except for AS3, where possibly the load balancing algorithm is not performing well when such a large number of nodes has to be classified. Finally, all three assignments have very similar throughput in all cases for token passing, whereas AS1 (random channel assignment to individual packets) works better in BRS.




\subsection{Hotspot Traffic}
We next discuss the results shown in the middle plots of Fig.~\ref{boxplotBrs} and Fig.~\ref{boxplotToken}, which illustrate the impact of uneven spatial injection distribution on performance. We remind that low/high values of $\sigma$ mean that traffic is hotspot/evenly distributed \cite{soteriou2006statistical}.



\vspace{0.1cm}
\noindent
\textbf{Latency.} In BRS, the hotspot behavior of traffic does not seem to have a large influence on the performance of the different assignment methods. The outlier, third quartile, and maximum values within the distribution seem to be mildly impacted by the hotspot nature of traffic. In general, BRS is resilient to such variations and actually could benefit from having a lower amount of nodes contending for the available channels. Still, the results show a small tendency to worse results when traffic is concentrated around a few nodes, possibly because of the nodes with higher load reaching higher backoff values. In AS3, this situation is avoided by proactively placing high-load nodes in different channels. Similarly, in token passing, latency is affected by the concentration of traffic around a given set of nodes mostly because the different assignment methods are able to provide tokens quickly to nodes that need it, even if they are spaced apart within the ring. This is clearly visible in the extreme case of $\sigma=0.05$. Similarly, outlier values seem to be larger when traffic is more hotspot. We also observe how AS2 fails to provide a good performance at low loads, and this behavior is exacerbated for very hotspot traffic. 

\vspace{0.1cm}
\noindent
\textbf{Throughput.} The throughput of BRS in its different implementations does not vary significantly with the type of spatial distribution of traffic, except for AS3, where a higher concentration of traffic around a few nodes seems to have a positive effect on the throughput. One reason could be that the most active nodes are distributed over the different channels so that contention is minimized. 
That does not happen in other assignment methods. 
Different behavior is observed in token passing, where the hotspot behavior of traffic modifies the throughput of the different assignment methods, with AS3 being affected a bit less. This is because if the load is concentrated around a small set of nodes, a large portion of the airtime is wasted while passing the token among these nodes. 


\subsection{Bursty Traffic}
Finally, we present the latency and throughput results for an increasingly bursty traffic. The results are shown in the right plots of Fig.~\ref{boxplotBrs} and Fig.~\ref{boxplotToken} for BRS and token passing, respectively. Temporal injection of traffic is modeled through the Hurst exponent \cite{soteriou2006statistical}, with higher values indicating more bursty behavior, i.e. longer bursts followed by longer silences. 

\vspace{0.1cm}
\noindent
\textbf{Latency.}
In BRS, it can be seen that the higher the value of $H$, the higher the latency in average and also the more unpredictable. This is because with an $H$ of 0.5, the packets are injected following a random Poisson process, which minimizes the probability of collisions. However, when increasingly bursty traffic is considered, the probability of packets being injected (and nodes trying to transmit) in the same exact cycle increases. The effect is multiplicative with the burstiness, as the effect of cascading collisions leads to an exponential increase of the backoff time. This affects the system at all loads.
On the other hand, token passing also suffers when bursty traffic is served, leading to very high latency especially for high values of $H$. The latency is a bit more stable than in the case of BRS, mainly because the protocol does not react with exponential backoffs, but rather with linear token passings to bursts of traffic. Still, the latency is much higher than that of BRS, discouraging its use for large number of nodes.

\vspace{0.1cm}
\noindent
\textbf{Throughput.} On one hand, it can be verified that in BRS, the saturation throughput remains rather constant across all assignments regardless of the value of the Hurst exponent. A possible reason could stem from the behavior of the backoff mechanism; bursty traffic leads to a large number of collisions which increases latency even for low loads, but the protocol may converge to a large backoff value that can accommodate the load even if it comes in bursts. In other works, the backoff mechanisms spreads out the bursts of traffic over time, until all nodes are backlogged.
On the other hand, it can be seen that in the case of token passing, the saturation throughput seems to drop significantly for higher numbers of $H$, to a point that the achieved throughput becomes comparable with that of BRS. A potential reason for this behavior is the lack of an adaptive mechanism to react to bursts; the token has to still move around the ring even if bursts of traffic lead to the generation of multiple packets in a given node, leading to gaps where the wireless channel remains silent. When traffic is less bursty, the probability of such events is lower. 

\section{Discussion}
\label{sec:discussion}
\begin{figure}[!t] 
\centering
\includegraphics[width=\columnwidth]{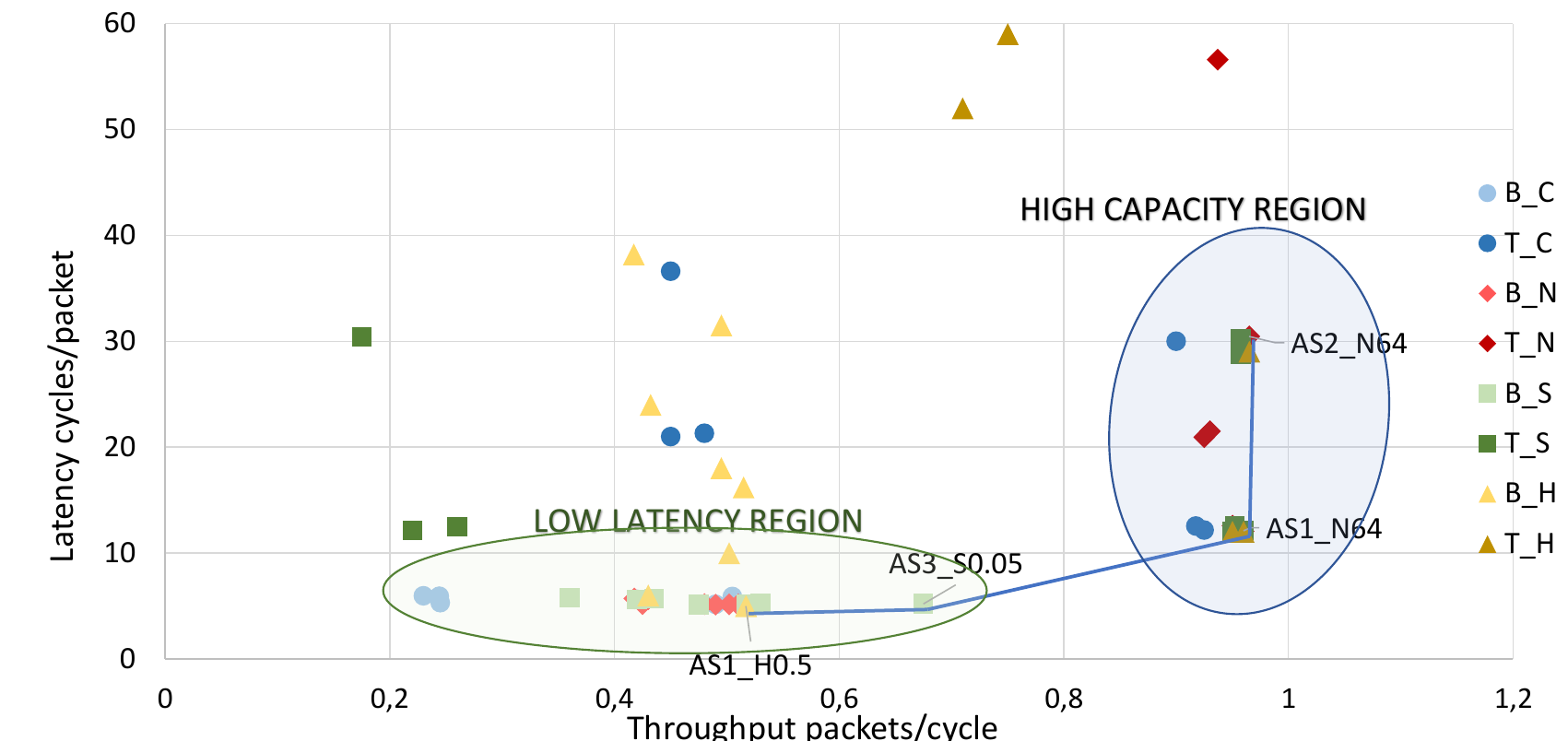}
\vspace{-0.5cm}
\caption{Summary of the latency and throughput results over all the protocols, assignment methods, and traffic conditions. $B$ and $T$ stand for BRS (random access) and token passing, $C$ and $N$ denote number of channels and nodes, whereas $S$ and $H$ represent the different spatial and temporal injection distributions. For instance, the $B$\_$C$ symbols represent the latency-throughput of all the assignment methods for BRS for different number of channels. Two desirable design spaces and a Pareto frontier are also given.}
\label{summaryResults}
\end{figure}

Figure \ref{summaryResults} plots the performance of all the compared protocols and assignments representing the zero-load latency (X axis) and saturation throughput (Y axis) of a particular protocol for a given number of channels and assignment method. 

In general terms, BRS is preferred over token in terms of zero-load latency given its ability to transmit immediately when the channels are idle. Hence, we see most BRS points located in a \emph{low latency region}. Among the assignment techniques, AS1 achieves similar results than AS3 and would probably be preferred as it does not require prior knowledge of the load of each node to assign the channels. On the downside, the throughput is half of that of token passing, at most. 

On the other hand, token passing can reach high throughput levels in the \emph{high capacity region}, close to the maximum total bandwidth of the wireless network. However, while putting more channels reduces the latency significantly, the best latency in token passing is still several cycles away from the BRS values. Finally, we observe that it is hard to provide a good channel assignment overall: AS3 requires prior knowledge on the traffic distribution, AS1 does not perform well for hotspot traffic and AS2 has high latency.

\section{Conclusions}
\label{sec:conclusions}
This paper has explored several techniques to extend random access and token passing MAC protocols to multiple channels for wireless chip-scale networks. In general, more channels alleviate the problems of both types of protocols, increasing the throughput of random access and cutting down the latency of token passing to a few tens of cycles. Additionally, random access is more resilient to hotspot and bursty traffic and more scalable to massive chip-scale networks. However, the higher throughput achievable with token renders the decision of the protocol (and assignment) to choose extremely challenging. Hence, we see a trend similar to that of single-channel protocols: it would be desirable to develop a multi-channel protocol that is able to seamlessly obtain the best of both paradigms. This will be explored in future work.



\end{document}